\documentclass{ws-ijmpa}
\usepackage[super,compress]{cite}

\usepackage{amsmath,amssymb,amsfonts,latexsym}
\usepackage{graphicx}

\def\noi{\noindent}
\newcommand{\foom}[1]{\protect\footnotemark[#1]}

\def\cm{\hspace*{1cm}}
\def\inch{\hspace*{1in}}

\def\Jl#1#2{{\it #1\/} {\bf #2},\ }

\def\ApJ#1 {\Jl{Astroph. J.}{#1}}
\def\CQG#1 {\Jl{Class. Quantum Grav.}{#1}}
\def\DAN#1 {\Jl{Dokl. AN SSSR}{#1}}
\def\GC#1 {\Jl{Grav. Cosmol.}{#1}}
\def\GRG#1 {\Jl{Gen. Rel. Grav.}{#1}}
\def\JETF#1 {\Jl{Zh. Eksp. Teor. Fiz.}{#1}}
\def\JETP#1 {\Jl{Sov. Phys. JETP}{#1}}
\def\JHEP#1 {\Jl{JHEP}{#1}}
\def\JMP#1 {\Jl{J. Math. Phys.}{#1}}
\def\NCB#1 {\Jl{Nuovo Cim. B}{#1}}
\def\NPB#1 {\Jl{Nucl. Phys. B}{#1}}
\def\NP#1 {\Jl{Nucl. Phys.}{#1}}
\def\PLA#1 {\Jl{Phys. Lett. A}{#1}}
\def\PLB#1 {\Jl{Phys. Lett. B}{#1}}
\def\PRD#1 {\Jl{Phys. Rev. D}{#1}}
\def\PRL#1 {\Jl{Phys. Rev. Lett.}{#1}}

\def\al{&}
\def\lal{&&{}}
\def\eq{Eq.\,}
\def\eqs{Eqs.\,}
\def\beq{\begin{equation}}
\def\eeq{\end{equation}}
\def\bear{\begin{eqnarray}}
\def\bearr{\begin{eqnarray} \lal}
\def\ear{\end{eqnarray}}
\def\earn{\nonumber \end{eqnarray}}
\def\nn{\nonumber\\ {}}

\def\nnn{\nonumber\\ \lal }
\def\nnnv{\nonumber\\[5pt] \lal }

\def\eql{\al =\al}

\def\tst{\textstyle}

\def\fract#1#2{{\tst\frac{#1}{#2}}}

\def\half{{\fract{1}{2}}}

\def\e{{\,\rm e}}

\def\diag{\mathop{\rm diag}\nolimits}

\def\const{{\rm const}}

\def\then{\ \Rightarrow\ }

\def\eqn#1{\eq\eqref{#1}}
\def\rf{\eqref}

\def\MN{^{\mu\nu}}
\def\mN{_\mu^\nu}

\def\R{{\mathbb R}}
\def\tT{{\widetilde T}}

\def\kappa{\varkappa}

\def\cy{cylindrical}
\def\cyl{cylindrically symmetric}

\def\wh{wormhole}
\def\whs{wormholes}
\def\asflat{asymptotically flat}
\def\intcon{integration constant}

\def\cy{cylindrical}
\def\cyl{cylindrically symmetric}

\begin{document}

\markboth{K.A. Bronnikov}
{String clouds and radiation flows as sources of gravity in static or rotating cylinders}


\title{String clouds and radiation flows as sources of gravity in static or rotating cylinders}

\author{K. A. Bronnikov\foom 1}

\address{VNIIMS, 	Ozyornaya 46, Moscow 119361, Russia;\\
	Institute of Gravitation and Cosmology, Peoples' Friendship University of Russia\\
	(RUDN University), ul. Miklukho-Maklaya 6, Moscow 117198, Russia;\\
	National Research Nuclear University ``MEPhI''
		(Moscow Engineering Physics Institute),\\
		Kashirskoe sh. 31, Moscow 115409, Russia\\
		kb20@yandex.ru		}
		
\maketitle


\begin{abstract}
   Static and stationary cylindrically symmetric space-times in general relativity are considered, 
   supported by distributions of cosmic strings stretched in the azimuthal ($\varphi$), longitudinal 
   ($z$) or radial ($x$) directions or and by pairs of mutually opposite radiation flows in any of 
   these directions. For such systems, exact solutions are obtained and briefly discussed, 
   except for radial strings (a stationary solution for them is not found); it is shown that static 
   solutions with $z$- and $\varphi$-directed radiation flows do not exist while for $z$-directed 
   strings a solution is only possible with negative energy density. Almost all solutions under 
   discussion contain singularities, and all stationary solutions have regions with closed timelike 
   curves, hence only their well-behaved regions admit application to real physical situations. 
\end{abstract}


\keywords{General relativity, cylindrical symmetry, rotation, exact solutions, cosmic strings,
                    radiation flows}  

\ccode{PACS numbers: 04.20.-q, 04.20.Jb, 04.40.Nr, 04.20.Gz}

\section{Introduction}

  Cylindrically symmetric space-times have attracted the researchers' attention soon after the advent 
  of general relativity (GR) as examples of the simplest geometries describing extreme deviations from 
  spherical symmetry and sources likes jets and flows, rods and (as we know now) cosmic strings.   
  The story began with the Levi-Civita static vacuum solution \cite{LC} (1919) and its stationary 
  counterparts with rotation, \cite{lanczos, lewis} and new solutions are still being found and studied. 
  The relative simplicity of cylindrical symmetry as compared to more general axial symmetry is 
  favorable for studying rotation effects in GR.  Among sources of gravity in known stationary exact 
  solutions of GR one can mention the cosmological constant  $\Lambda$, \cite{vac-Lam1, vac-Lam2}
  scalar fields with or without self-interaction potentials,\cite{BLem13,BK15,erices} rigidly or differentially 
  rotating dust, \cite{van_stockum,bonnor09} electrically charged dust, \cite{iva02b} dust combined with 
  a scalar field \cite{santos82}, perfect and anisotropic fluids  
  \cite{hoen79, davids00, skla99, iva02a, anis1, anis3} etc, see also the reviews 
  \cite{exact-book, BSW19} and references therein. 
  
  We adhere to GR as a perfectly successful theory on scales, say, from centimeters 
  to parsecs, while its diverse extensions are mostly designed for much smaller or
  much larger scales. In addition, solutions of many alternative theories can be obtained 
  from those of GR by various solution generation methods like conformal mapping between
  the Einstein and Jordan frames in $f(R)$ and scalar-tensor theories. 

  In our previous papers, \cite{BLem09, BLem13, BK18, BBS19} some static and stationary 
  \cy\ \wh\ solutions were found and discussed, and a construction of \asflat\ \wh\ configuration
  was suggested that does not contain exotic matter violating the Weak Energy Condition.
  The latter required, as a source of gravity, either a special kind of anisotropic fluid \cite{BK18}
  or a stiff perfect fluid. \cite{BBS19}.  This paper is devoted to a search for other sources 
  of gravity possibly providing \wh\ solutions for stationary \cyl\ space-times. Specifically, we  
  consider distributions of cosmic strings stretched in the azimuthal, longitudinal or radial directions,
  and mutually opposite radiation flows in any of these directions. By the structure of their 
  stress-energy tensors (SETs) $T\mN$, they represent special examples of anisotropic fluids. 
  We follow the formalism described in Ref. \refcite{we19} which includes splitting of the 
  Ricci ($R\mN$) and Einstein ($G\mN$) tensors into static and rotational parts \cite{BLem13} 
  and using the harmonic radial coordinate \cite{kb79} for \cy\ metrics.
  We enumerate possible exact solutions with such sources and briefly discuss their basic properties.
  
\section{Basic relations}   

Consider a stationary \cyl\ metric
\bearr                                                    \label{ds-rot}
         ds^2 = \e^{2\gamma(x)}[ dt - E(x)\e^{-2\gamma(x)}\, d\varphi ]^2- \e^{2\alpha(x)}dx^2 
	- \e^{2\mu(x)}dz^2 - \e^{2\beta(x)}d\varphi^2,
\ear
  where $x$, $z\in \R$, $\varphi\in [0, 2\pi)$ are the radial, longitudinal and angular coordinates,
  respectively. The coordinate $x$ is specified up to a change $x \to f(x)$, and its range depends 
  on both its choice and on the geometry itself. The only off-diagonal metric component $E$ 
  describes rotation, with the vorticity $\omega(x)$ (in terms of an arbitrary coordinate $x$)
  given by \cite{BLem13, kr2, kr4} $\omega(x) = \half (E\e^{-2\gamma})' \e^{\gamma-\beta-\alpha}$
  (the prime denotes $d/dx$). Next, choosing the reference frame comoving to matter (as it moves  
  along $\varphi$), we require $T^3_0 = 0$, then the Einstein equations lead to 
  $R_0^3 \sim (\omega \e^{2\gamma+\mu})' = 0$, so that 
\beq       	      					\label{omega}
	\omega = \omega_0 \e^{-\mu-2\gamma}
		\qquad  {\rm and} \qquad
	E(x) = 2\omega_0 \e^{2\gamma(x)} \int \e^{\alpha+\beta-\mu-3\gamma}dx,
\eeq
  with $\omega_0 = \const$. The nontrivial components of the Einstein equations 
  $R\mN\! =\! - \kappa \tT\mN\\ \equiv -\kappa(T\mN - \half \delta\mN T^\alpha_\alpha)$ and the 
  constraint equation $G^1_1 = - \kappa \tT^1_1$ may be written as 
\bear                 \label{EE}
      R^0_0 \eql -\e^{-2\alpha}[\gamma'' + \gamma'(\sigma' -\alpha')] - 2\omega^2 = -\kappa \tT^0_0,
\nn     
      R^2_2 \eql -\e^{-2\alpha}[\mu'' + \mu'(\sigma' -\alpha')] = - \kappa \tT^2_2,  
\nn   
      R^3_3 \eql -\e^{-2\alpha}[\beta'' + \beta'(\sigma' -\alpha')] + 2\omega^2 = - \kappa \tT^3_3,  
\nn
      R^0_3 \eql G^0_3 =  E \e^{-2\gamma}(R^3_3 - R^0_0) = - \kappa E \e^{-2\gamma}(T^3_3 - T^0_0), 
\nn
	G^1_1 \eql \e^{-2\alpha} (\beta'\gamma'  + \beta'\mu' + \gamma' \mu') + \omega^2 = - \kappa T^1_1.      
\ear
 where $\sigma = \beta + \gamma + \mu$ and $\kappa = 8\pi G$ is the gravitational constant.
  As follows from \rf{EE}, the diagonal components $R\mN$ and $G\mN$ split into static parts
  ${}_s R\mN$ and ${}_s G\mN$ (those for the metric \rf{ds-rot} with $E=0$)  plus contributions 
  from $\omega$ \cite{BLem13}, in particular,  
\beq    		\label{G-omega}  
		G\mN = {}_s G\mN + {}_\omega G\mN,  \qquad
   					{}_\omega G\mN = \omega^2 \diag (-3, 1, -1, 1),  
\eeq
  Each of the tensors ${}_s G\mN$ and ${}_\omega G\mN$ separately obeys the conservation
  law $\nabla_\alpha G^\alpha_\mu =0$ according to this static metric. Thus 
  ${}_\omega G\mN/\kappa$ acts as an  additional SET of a ``vortex gravitational field'' with 
  a negative effective energy density $T^0_0 = -3\omega^2/\kappa$, favorable for \wh\
  existence, as is confirmed by a few examples.\cite{BLem13,BK15,kr4,BBS19}.
  
  What is important, it is sufficient to solve the diagonal components of the Einstein equations, 
  their single off-diagonal component then holds as well \cite{BLem13}.  

\medskip\noi
  {\bf Anisotropic fluids.} Consider\cite{anis1,anis3, we19} nondissipative 
  matter with energy density $\rho$ and three principal pressures $p_i$ in mutually orthogonal 
  directions in a comoving reference frame, so that the SET has the form\footnote
  	{Tetrad indices are written in parentheses.} 
  $T^{(\alpha\beta)} = \diag (\rho, p_1, p_2, p_3)$  in an orthonormal tetrad 
  $\big( e^\mu_{(\alpha)} \big) = \big( u^\mu, \phi^\mu, \chi^\mu, \psi^\mu \big)$, 
  where $u^\mu$ is the fluid 4-velocity ($u_\mu u^\mu =1$)  while the vectors 
  $\phi^\mu, \chi^\mu, \psi^\mu$ are spacelike. Choosing (arbitrarily) one of them as 
  $\phi^\mu = e^\mu_{(1)}$, we have
   $g\MN  =  u^\mu u^\nu  - \phi^\mu \phi^\nu - \chi^\mu \chi^\nu - \psi^\mu \psi^\nu$ and   
\beq             \label{SET3}
	T\MN = (\rho + p_1) u^\mu u^\nu  - p_1 g\MN 
			+ (p_2 - p_1)\chi^\mu \chi^\nu + (p_3 - p_1)\psi^\mu \psi^\nu. 
\eeq   

 Consider now this kind of matter in  the metric \rf{ds-rot}, assuming that $p_1= p_x$ is the radial 
 pressure,  $p_2 = p_z$ and $p_3 = p_\varphi$ being pressures in the $z$ and $\varphi$ directions.
 Then one can verify that the conservation law  $\nabla_\nu T\mN =0$ does not contain $E$ and takes 
 the same form for rotating and nonrotating fluids:
\beq 				\label{cons}
		p'_x + (\rho + p_x) \gamma' + (p_x - p_z) \mu' + (p_x - p_\varphi) \beta' =0. 
\eeq    
  
  Let us now assume in \rf{SET3} the equations of state
\beq    \label{w_i}
	   p_x = w_1 \rho, \qquad p_z = w_2 \rho, \qquad p_\varphi = w_3 \rho,  \qquad w_i = \const,
\eeq    
  and choose the harmonic radial coordinate defined by the condition \cite{kb79}
  $\alpha = \beta + \gamma + \mu$. Then three diagonal equations \rf{EE} and the constraint read 
\bearr         			\label{EE00}
		\e^{-2\alpha} \gamma'' = - 2 \omega^2 + \half \kappa\rho (1 + w_1 + w_2 + w_3),
\\ \lal         			\label{EE22}
		\e^{-2\alpha} \mu''  =  \half \kappa\rho (-1 + w_1 - w_2 + w_3),
\\ \lal          			\label{EE33}								
		\e^{-2\alpha} \beta'' =  2 \omega^2 + \half \kappa\rho (- 1 + w_1 + w_2 - w_3),
\\ \lal                  \label{int}
	\e^{-2\alpha}(\beta'\gamma'\! + \beta'\mu' \! + \gamma' \mu') =  - \omega^2 + w_1\kappa \rho.  
\ear            

\section{String clouds}

  Let us now discuss special cases of \eqs \rf{EE00}--\rf{int} corresponding to distributions of 
  cosmic strings stretched in one of the directions $\phi^\mu, \chi^\mu, \psi^\mu$, so that the 
  factors $(w_1, w_2, w_3)$ in the expressions \rf{w_i} take the forms $(-1,0,0)$, $(0,-1,0)$ 
  or $(0,0,-1)$. In all cases there are five unknowns: the metric functions 
  $\beta(x), \gamma(x), \mu(x), E(x)$ and the  density $\rho(x)$.

\medskip\noi
{\bf Azimuthal strings: $(w_1, w_2, w_3) = (0,0,-1)$.}   
  For such distributions of coaxial circular string loops, the stationary solution for the metric 
  reads \cite{we19}
\beq                     \label{ds--1}
		ds^2 = 2|\omega_0|r_0 x \bigg[ dt + \frac {1- E_0 x}{2\omega_0 x}d\varphi\bigg]^2
				- r_0^2 \e^{2\mu} dx^2 - \e^{2\mu} dz^2 - \frac{r_0}{2|\omega_0|x }d\varphi^2.		
\eeq
  where $r_0=\const$ is an arbitrary length scale, and $E_0$ is an \intcon\ from \rf{omega}.   
  The unknowns $\rho(x)$ and $\mu(x)$ are connected by the relation due to \eqn{EE22},
\beq                                         \label{rho1}
		r_0^2 \kappa\rho  = - \mu'' \e^{-2\mu},	
\eeq    
  thus the density distribution in the string cloud remains arbitrary.  The behavior of 
  $g_{00} = \e^{2\gamma}$ indicates singularities on both ends, $x \to 0$ and $x\to\infty$. 
  Furthermore, the metric component $g_{33}$ is 
\beq
		g_{33} = - \e^{2\beta} + \e^{-2\gamma} E^2 = \frac {r_0 E_0(E_0 x -2)}{2|\omega_0|}.
\eeq          
  It means that if $E_0>0$, $g_{33} < 0$, and the spatial section $t=\const$ of the manifold 
  \rf{ds--1} has a normal  signature only in the range $0 < x < 2/E_0$, while at $x > 2/E_0$ we have 
  $g_{33} > 0$, the closed coordinate lines of $\varphi$ are timelike, violating the causality principle.
  If $E_0 < 0$, $g_{33} > 0$ in the whole range $x \in \R_+$. If $E_0 =0$, then $g_{33} \equiv 0$,
  making the spatial sections $t = \const$ degenerate. The metric as a whole is certainly not 
  degenerate, and other spatial sections can have a normal signature.  

  The corresponding {\it static solution\/} is obtained from the same equations with 
  $\omega_0 =0$. They  lead to $\beta' = \gamma' =0$ and the same relation \rf{rho1} for $\mu$ 
  and $\rho$. Without  loss of generality, the metric can be written as
\beq
		ds^2 = dt^2 - \e^{2\mu (x)} (r_0^2 dx^2 + dz^2) - r_0^2 d\varphi^2.
\eeq    
  It describes a congruence of cylinders of equal radii but an $x$-dependent scale along the $z$ 
  axis, with a distribution of closed strings as $\varphi$-circles with an $x$-dependent density profile.

\medskip\noi
{\bf  Longitudinal strings: $(w_1, w_2, w_3) = (0,-1,0)$.}   
  For such strings stretched along the $z$ axis, the conservation law \rf{cons} 
  leads to $\gamma' = - \mu'$, hence $\mu'' = - \gamma''$. Comparing the expressions for 
  $\mu''$ from \rf{EE22} and for $\gamma''$ from \rf{EE00}, we see that 
\beq                                   \label{rho3} 
	     \mu'' = - \kappa \rho\e^{2\alpha} = - \gamma'' = 2\omega_0^2 \e^{2\beta-2\gamma},       
\eeq    
  we see that $\rho < 0$. Thus we inevitably obtain a negative energy density and do not 
  consider this case any more.
  
  For static systems we again obtain \rf{rho3} but now with $\omega_0 =0$, hence 
  $\rho  \equiv 0$, vacuum. Thus static \cyl\ distributions of longitudinal cosmic strings do not exist, 
  while stationary ones require negative energy density.

\medskip\noi
{\bf Radial strings: $(w_1, w_2, w_3) = (-1,0,0)$.}   
  It is a cloud of radial strings which, imagined around a symmetry axis in flat space, 
  would resemble a barbed wire. In curved space there can be a geometry without a symmetry 
  axis, e.g., that of a wormhole. Let us consider the corresponding equations.
  
  The conservation law \rf{cons} leads to $\rho = \rho_0 \e^{-\mu -\beta}$. Using it in 
  \eqs \rf{EE00}--\rf{int} leads to $\mu'' = \beta'' + \gamma''$ 
  (hence we can put $\mu = \beta + \gamma + mx$, $m=\const$) and 
\bearr                        \label{00-5}
		\gamma'' = -2\omega_0^2 \e^{2\beta - 2\gamma},
\\ \lal 			\label{33-5}
		\beta'' = 2\omega_0^2 \e^{2\beta - 2\gamma} - \kappa \rho_0 \e^{2\gamma + \mu + \beta}, 		
\\ \lal             \label{int5}
	        \beta' \gamma' + (\beta'+\gamma') (\beta'+\gamma' + m)
	        	=  -\omega_0^2 \e^{2\beta - 2\gamma} - \kappa \rho_0 \e^{2\gamma + \mu + \beta}, 		
\ear  
   It is hard to solve these equations analytically, but some qualitative observations can be made.  
  In particular, wormhole solutions (implying a regular minimum of $\beta(x)$) are not forbidden 
  in general, but if $m=0$, any regular extremum of $\beta(x)$ or $\gamma(x)$ is      
  incompatible with $\rho > 0$ due to negativeness of the r.h.s. in \rf{int5}. Wormholes, if any, 
  require $m \ne 0$ and are asymmetric under the change $x \to -x$.
  
\medskip
  {\it Static solutions\/} ($\omega_0=0$) are easily found from \rf{00-5}--\rf{int5}: 
  we have $\gamma'' =0 \then \gamma = hx$ ($h= \const$, a $t$-scale is chosen), and $\beta$ is 
  derived from \rf{int5}.  The resulting solutions consist of two branches. In the first one, 
  the metric and the density are  
\bearr                 \label{ds51} 
	ds^2 = \e^{2hx} dt^2 - \frac{k^4}{\kappa \rho_0^2}\frac{\e^{-2hx}}{\cosh^4 kx}dx^2
	                - \frac{k^2 \e^{-2hx}}{\cosh^2 kx}
	                		\bigg( \frac{\e^{2mx}}{r_0} dz^2 + r_0 \e^{-2mx} d\varphi^2 \bigg),
\nnn
	  \rho = \kappa \rho_0^2 \e^{2hx} \frac{\cosh^2 kx}{k^2},\qquad
	  k = \sqrt{h(h+2m)} \geq 0, \qquad    h(h+2m) \geq 0.	  
\ear	   	                		
  In the second branch, 
\bearr  			\label{ds52}
		ds^2 = \e^{2hx} dt^2 - r_0^2 \e^{-2hx} dx^2 - \e^{2(m-h)x} dz^2 
										- r_0^2 \e^{-2(m+h)x} d\varphi^2,
\nnnv
		\rho = (\rho_0/r_0) \e^{2hx}, \qquad       \kappa \rho_0 r_0 = h(h+2m).  
\ear      
   In both branches the range of $x$ is $x \in \R$. The metrics \rf{ds51} and \rf{ds52} do not 
   describe \whs\ at any values of the parameters; the absence of a minimum of $\beta(x)$ is 
   already clear from \eqn{33-5} since $\beta'' < 0$ if $\omega_0 =0$. A regular axis 
   corresponding to $\e^\beta =0$ is also impossible, which is natural since, in non-\wh\  
   space-time, radial strings must begin from a source on a symmetry axis.

\section{Radiation flows}

  A null radiation flow with intensity $\Phi$ has the SET $T\mN = \Phi k_\mu k^\nu$, where
  $k^\mu$ is a null vector in a given spatial direction. If there are two such flows of 
  equal intensity in mutually opposite directions, say, $\pm x^1$, their SETs add into 
  $T\mN = \rho \diag(1, -1, 0, 0)$ ($\rho = 2\Phi$), describing a special case of anisotropic 
  fluid.\cite{we19,we-kim16}  Let us consider such flows in different directions as sources 
  of gravity in \eqs \rf{EE00}--\rf{int}. 

\medskip\noi
{\bf Asimuthal flows: $(w_1, w_2, w_3) = (0,0,1)$.}   
  The conservation law leads to $\beta'=\gamma'$, while \eqs \rf{EE00}--\rf{EE33} yield
  $\beta''+\gamma'' = \mu'' =0$, and without loss of generality we have
\beq
	\beta = b x + \ln r_0, \qquad \gamma =b x, 
	\qquad \mu = m x, \qquad	m = - \frac {b^2 + \omega_0^2 r_0^2}{2 b},
\eeq    
  where $b = \const$, and the last equality follows from \rf{int}. The integral in \rf{omega} 
  is easily found, while $\rho$ is obtained from \rf{EE00}, so that finally
\bearr
		ds^2 = \e^{2 bx} [dt -2\omega_0 r_0^2 (E_0+x) d\varphi]^2
		           -r_0^2 \e^{2ax} dx^2 - \e^{2mx} dz^2 - r_0^2 \e^{2bx} d\varphi^2,
\nnnv
		\kappa\rho =2\omega_0^2 \e^{-2ax}, \qquad   
		a =  \frac {3 b^2 - \omega_0^2 r_0^2}{2 b},  \qquad   E_0 = \const.
\ear    
   The solution is defined for $x \in \R$. At $a =0$ (that is, $3b^2 = \omega_0^2 r_0^2$), we 
   obtain $\rho = \const$, a homogeneous density. In other cases $\rho \to \infty$ as either
   $x \to \infty$ or $x \to - \infty$. If (say) $b>0$, $x\to \infty$ is spatial infinity and $x\to -\infty$
   the axis. A feature of interest is that if $3b^2 > \omega_0^2 r_0^2$ (that is, $a > 0$), then
   $\rho$ infinitely grows at spatial infinity. Lastly, a calculation of $g_{33}$ shows that 
   $\varphi$-circles are inevitably timelike at large $x$, e.g., at $|x| > 1/(2|\omega_0|r_0)$
   if $E_0 = 0$.
   
   In the static case $\omega_0 = 0$, we obtain $\rho=0$, hence the vacuum solution.\cite{LC}.        

\medskip\noi
{\bf Longitudinal flows: $(w_1, w_2, w_3) = (0,1,0)$.}   
  For this case, the solution is already known \cite{we19} and consists of two branches, one 
  with a constant $b >0$, the other with $b=0$. The metric reads 
\bearr
	ds^2 = (\e^{2bx}-1)^{2/3}\Big[dt -2\omega_0(E_0 - b \coth bx) d\varphi\Big]^2 
	- r_0^2 b^{4/3} \e^{2bx} dx^2 
\nnn \inch\cm	
	- (\e^{2bx}-1)^{2/3} dz^2 - r_0^2 \bigg[\frac {b^2 \e^{bx}}{\sinh bx}\bigg]^{2/3}d\varphi^2,
\ear
   for $ b>0$, while for $b =0$ we have 
\beq
		ds^2 = x^{2/3} \Big[dt - \frac{2\omega_0}{x}(E_0x -1) d\varphi\Big]^2
		- r_0^2 dx^2 - x^{2/3} dz^2 - r_0^2 x^{4/3} d\varphi^2 
\eeq         
  where $E_0 = \const$. For the density we obtain 
\beq
		\kappa\rho r_0^2 = \frac{8b^2}{3(\e^{2bx}-1)^2}\ \ {\rm for} \ \ b>0,
				\qquad
		\kappa\rho r_0^2 = 2/ (3x^2) \ \ {\rm for} \ \ b=0.
\eeq    
  The solutions are defined for $x \in \R_+$. In all cases the value $x=0$ corresponds to 
  radii $r = \e^\beta \to \infty$ which is a singularity due to $\rho \to \infty$. Other features
  depend on $b$ and $E_0$. Closed timelike curves are observed at large $x$ if $b =0$, 
  $E_0 \ne 0$. 
  
  In the static case, the equations lead again to $\rho =0$, the vacuum solution. 

\medskip\noi
{\bf Radial flows: $(w_1, w_2, w_3) = (1,0,0)$.}   
  The conservation law \rf{cons} leads to $\rho = \rho_0 \e^{-\beta - 2\gamma - \mu}$,
  $\rho_0 = \const$. Equation \rf{EE22} reads $\mu'' =0 \then  \mu = mx$, $m = \const$.
  For $\gamma(x)$ and $\beta(x)$ we then obtain the equations
\bearr                     \label{EE6}
		\gamma'' = - 2 \omega_0^2 \e^{2\beta - 2\gamma} + \kappa \rho_0 \e^{\beta + mx},
\nnn		
		\beta'' = 2 \omega_0^2 \e^{2\beta - 2\gamma}. 
\ear      
  It is hard to find their general solution, but a special one can be found by assuming   
  $\e^{\beta-2\gamma} = \e^{mx}/r_0$, $r_0 = \const$. With this ansatz \eqs \rf{EE6} lead to
\beq
		\kappa \rho_0 = 3h^2 \omega_0^2, \qquad  {\rm and} \qquad
		\beta''-\gamma'' = \omega_0^2 \e^{2\beta - 2 \gamma}
\eeq
  which is easily solved. Substituting the solution to \rf{omega} to find $E(x)$ and taking 
  into account the constraint \rf{int}, we finally obtain for $m \ne 0$
\bearr                        \label{ds6}
 		ds^2 = \frac{k^2 \e^{-2mx}}{r_0^2 \omega_0^2 \sinh^2 kx}
 				\bigg[ dt+ \frac 2 {\omega_0}(k \coth kx - E_0) d\varphi \bigg]^2 				 		
\nnn \cm\cm
 		- \frac{k^6 \e^{-2mx}}{r_0^4 \omega_0^6 \sinh^6 kx} dx^2 - \e^{2mx} dz^2 
 		- \frac{k^4 \e^{-2mx}}{r_0^2 \omega_0^4 \sinh^4 kx} d\varphi^2,
\ear   
   where $k = |m|/\sqrt{2}$ and $E_0$ are constants. For $m=0$ we arrive at
\beq                        \label{ds6'}
		  ds^2 = \frac 1 {r_0^2 \omega_0^2 x^2}
		  			\bigg[ dt + \frac 2{\omega_0 x}(1-E_0 x) d\varphi\bigg]^2
       - \frac {dx^2} {r_0^4 \omega_0^6 x^6} - dz^2 - \frac {d\varphi^2}{r_0^2 \omega_0^4 x^4};		
\eeq      
  the range of $x$ in both \rf{ds6} and \rf{ds6'} is $x > 0$. For the density we have
\beq
		\rho = \frac {\rho_0}{k^4}\omega_0^4 r_0^3 \e^{2mx} \sinh^4 kx  \ \ \ {\rm for} \ \ m\ne 0,
		\qquad 
		\rho = \rho_0 \omega_0^4 r_0^3 x^4 \ \ \ {\rm for} \ \ m= 0.
\eeq    
  In all cases there exist timelike $\varphi$-circles at sufficiently large $x$, and $x\to \infty$
  is a singularity where $\e^\beta \to 0$ and $\rho \to \infty$.
            
  For {\it static configurations} ($\omega_0 =0$) we have  $\rho = \rho_0 \e^{-\beta - 2\gamma - \mu}$
  and $\mu''\! =\! \beta'' \!=\!0 \then  \mu= mx,\ \beta = bx + \ln r_0$, $m, b , r_0 = \const$. From
  \rf {int} we obtain an expression for $\gamma(x)$, and the solution takes the form
\bearr  \label{ds6-0} 
		ds^2 = \e^{2\gamma} dt^2 - r_0^2 \e^{2\gamma +2(m+b)x} dx^2
		                   -  \e^{2mx} dz^2 - r_0^2 \e^{2 bx} d\varphi^2,
\nnn
	\rho = \frac {\rho_0}{r_0} \e^{-2\gamma - (b+m)x}, \cm
	\e^\gamma =  \exp\bigg[\frac{\kappa \rho_0}{(m+b)^2} \e^{(m+b)x} - \frac{mbx}{m+b}\bigg],
\ear  
  where $b+m \ne 0$. The properties of this simple solution depend on the values of $b$ and $m$
  and are evident from the expressions \rf{ds6-0}.
  
\section{Concluding remarks}

  {\bf 1.} Among new exact solutions obtained here, with such sources of gravity as clouds of 
  cosmic strings and pairs of mutually opposite radiation flows, there are no nonsingular 
  ones, and stationary solutions contain regions with closed timelike curves. It evidently 
  means that not these global solutions but only their physically acceptable regions can be
  used in modeling some real situations in the nature. 

\smallskip\noi   
  {\bf 2.} It has been shown that stationary solutions with $z$-directed strings require $\rho < 0$,
  while static solutions with $z-$ or $\varphi-$directed radiation flows do not exist. On the 
  contrary,  $\varphi-$directed cosmic strings admit arbitrary density distributions.
    
\smallskip\noi   
  {\bf 3.} String distributions and radiation flows in $z$ and $\varphi$ directions possess zero radial 
  pressure $p_x$, making it possible to match the corresponding solutions to external vacuum 
  ones, with or without rotation, on any hypersurface $x = \const$, without need for surface 
  densities and tensions.

\smallskip\noi   
  {\bf 4.} Among the exact solutions under consideration, there are no wormhole ones. Throats as
  minima of $r (x) = \e^{\beta(x)}$ are possible in the cases of \eqs \rf{00-5}--\rf{int5} and 
  \rf{EE6} (to be solved numerically), and only solutions to \rf{EE6} can be symmetric 
  with respect to the throat. In the author's view, these cases deserve a separate study.  

\section*{Acknowledgments}

   I am grateful to Milena Skvortsova and Sergei Bolokhov for helpful discussions.
  This publication was supported by the RUDN University program 5-100 and by the 
  Russian Basic Research Foundation grant 19-02-00346. 
  The work was also performed within the framework of the Center FRPP 
  supported by MEPhI Academic Excellence Project (contract No. 02.a03.21.0005,
  27.08.2013).

\end{document}